\documentclass[11pt]{article}
\usepackage[final]{acl}
\usepackage{algorithm}
\usepackage{algpseudocode}
\usepackage{booktabs} 
\usepackage{multirow} 
\usepackage[table]{xcolor}
\usepackage{times}
\usepackage{latexsym}
\usepackage[T1]{fontenc}
\usepackage[utf8]{inputenc}
\usepackage{microtype}
\usepackage{inconsolata}
\usepackage{graphicx}

\hypersetup{colorlinks=true, linkcolor=blue, citecolor=blue, urlcolor=blue}

\begin{document}

\title{Not All Duplicates Are Coordination: Generic vs. Non-Generic Duplicate Campaigns in Information Operations}

\author{
  Ashfaq Ali Shafin\textsuperscript{1} \qquad
  Khandaker Mamun Ahmed\textsuperscript{2} \\
  \textsuperscript{1} Mathematics and Computer Science, Augustana College, USA \\
  \texttt{shafinashfaqali21@gmail.com} \\
  \textsuperscript{2} Beacom College of Computer and Cyber Sciences, Dakota State University, USA \\
  \texttt{khandakermamun.ahmed@dsu.edu}
}

\maketitle

\begin{abstract}
Duplicate content is widely used to study coordinated behavior in social media information operations (IOs), but not all repetition provides equally meaningful evidence of coordination. Generic, reusable, or low-information posts may create noisy account-account links when projected into coordination graphs. We study this problem using 187,000 English-language tweets from six Russian Twitter Information Operations datasets. We introduce a generic/non-generic distinction for duplicate campaigns, label tweets using an LLM-assisted protocol with independent human validation, and train supervised classifiers over sentence embeddings to scale the labels. We construct duplicate campaigns using lexical similarity and two embedding-based methods. Generic campaigns are rare under lexical matching but account for nearly 39\% of campaigns detected by embedding-based methods. Restricting graphs to non-generic campaigns reduces graph size and the largest connected component while increasing density, suggesting a smaller but more focused coordination structure. These findings show that duplicate-based coordination analysis should consider both textual similarity and semantic specificity.
\end{abstract}

\section{Introduction}

Social media platforms are used for political discussion, news sharing, and public mobilization, but they are also exploited by coordinated actors to amplify narratives, manipulate attention, and spread misleading or harmful content~\cite{RCHS23,ZCDSSB19,luceri2024unmasking, shafin2025toxicity, shafin2025language}. Detecting such behavior is difficult because researchers often lack privileged platform-side metadata, deleted account histories, or complete interaction traces. As a result, public content-based signals, including repeated or near-repeated text, remain important for studying possible coordination~\cite{pacheco2021uncovering,luceri2024unmasking,shafin2026iox}.

A common approach is to group repeated posts into \emph{duplicate campaigns} and connect accounts that participate in the same campaign. This projection can reveal groups of accounts repeatedly disseminating similar messages. However, duplicate content is not uniformly informative. Some repeated posts contain specific claims, targets, events, or narratives, while others are generic greetings, short praise, vague reactions, platform-generated updates, or reusable expressions. Treating these two forms of repetition as equivalent may create weak edges and distort the interpretation of account-account coordination graphs.

We study the distinction between \emph{generic} and \emph{non-generic} duplicate campaigns. We define semantic specificity as the extent to which a tweet anchors its meaning to identifiable claims, entities, events, policies, accusations, groups, or narratives. A post can be frequently repeated and still be non-generic if it carries a specific political or informational message. Conversely, a post can be fluent or emotionally expressive and still be generic if it can be reused across contexts without changing its meaning. Thus, broad slogans, greetings, motivational quotes, and low-information calls to action are treated as generic unless they are tied to a specific target, event, claim, or narrative frame.

We ask three research questions:

\noindent
\textbf{RQ1:} How common are generic duplicate campaigns in IO datasets?

\noindent
\textbf{RQ2:} How do generic and non-generic campaigns differ in size, account participation, textual specificity, and temporal behavior?

\noindent
\textbf{RQ3:} How does restricting duplicate-based coordination graphs to non-generic campaigns affect graph structure?

This paper makes three contributions. First, we operationalize a generic/non-generic distinction for duplicate campaigns and validate the annotation rubric with independent human verification. Second, we evaluate this distinction on 187,000 English-language tweets from six Russian Twitter Information Operations datasets using three duplicate-detection methods. Third, we show that generic campaigns can substantially inflate the apparent scale of coordination, especially for embedding-based methods, and that filtering to non-generic campaigns yields smaller but denser graphs centered on more narrative-specific repetition.

\section{Related Work}
\label{sec:related}

\paragraph{Coordination and duplicate-content analysis.}
Coordinated behavior on social media has been studied through shared behavioral traces, including URLs, hashtags, retweets, temporal synchronization, and content similarity \citep{pacheco2021uncovering,nizzoli2021coordinated,luceri2024unmasking,minici2025iohunter}. Textual repetition provides a particularly accessible signal because it can reveal accounts that disseminate identical or closely related messages using publicly observable content. Prior studies have used exact or near-duplicate text to identify coordinated groups and characterize their dissemination behavior \citep{PFM20,VNCCGL24}. More recent duplication-focused work has used repeated content to construct campaigns and account-level networks for studying misinformation and information operations \citep{shafin-carbunar-2026-duplicating,shafin2026iox}. These approaches demonstrate the utility of textual similarity for identifying coordinated dissemination, but they primarily determine whether posts are sufficiently similar to be grouped together. They do not explicitly model whether the repeated message is generic and broadly reusable or sufficiently specific to represent a shared claim, event, target, or narrative. Our work addresses this distinction by incorporating semantic specificity into duplicate-campaign analysis.

\paragraph{Noisy user-generated text and semantic similarity.}
User-generated social-media text poses distinct challenges for NLP because it is often short, informal, non-canonical, and lexically variable. Prior work has shown that normalization choices can affect downstream processing of noisy Twitter text \citep{van-der-goot-etal-2017-normalize}, while evaluations of text representations on Twitter demonstrate that improvements observed on standard text do not necessarily transfer uniformly to noisy user-generated content \citep{wang-etal-2020-empirical}. Recent work on sentence embeddings similarly highlights the importance of robustness to lexical variation and non-standard forms in user-generated content \citep{nishimwe-etal-2024-making}. These challenges are particularly relevant to duplicate detection, where semantically equivalent messages may differ substantially at the surface level. Early work on Twitter paraphrases showed that meaning-preserving messages can be expressed through different lexical forms \citep{xu-etal-2013-gathering}, and subsequent work demonstrated the prevalence and importance of lexically divergent paraphrases in short Twitter messages \citep{xu-etal-2014-extracting}. The SemEval Paraphrase and Semantic Similarity in Twitter task further formalized both binary paraphrase identification and graded semantic similarity for tweet pairs \citep{xu-etal-2015-semeval}. Large-scale paraphrase resources have also been constructed from Twitter by linking independently written tweets that refer to shared URLs \citep{lan-etal-2017-continuously}. Collectively, this literature motivates semantic approaches that can recover related messages beyond exact or near-exact lexical overlap. However, semantic similarity alone does not indicate how informative or context-specific the shared content is. Two posts may be highly similar while expressing a reusable greeting, slogan, reaction, or generic call to action rather than a specific narrative.

\paragraph{Semantic specificity and generic repetition.}
Related NLP research has examined variation in the amount and specificity of information expressed in text. \citet{li-nenkova-2015-detecting} study the identification of content-heavy sentences, distinguishing sentences that convey substantial informational content from those that contribute less content. \citet{li-etal-2016-improving} develop annotation guidelines for sentence specificity and show that specificity can be treated as a systematic linguistic property influenced by both the information expressed in a sentence and its discourse context. Our notion of semantic specificity is related to this literature but serves a different analytical purpose. We operationalize specificity for repeated social-media content by distinguishing generic messages that can be reused across contexts from non-generic messages anchored to identifiable claims, entities, events, policies, accusations, targets, groups, or narratives. This distinction complements prior work on semantic similarity: similarity determines whether messages express related content, whereas specificity characterizes how context-dependent and narratively informative that shared content is. We bring these two dimensions together to examine how generic and non-generic repetition is recovered by lexical and embedding-based duplicate-detection methods and how this distinction affects the account-account coordination graphs constructed from duplicate campaigns.

\section{Data and Methodology}
\label{sec:method}

\noindent\textbf{Data.}
We use six publicly released Twitter Information Operations datasets attributed to Russian information operations and published by Twitter between 2019 and 2021. We focus on Russian-attributed English-language IO data because these releases provide a public, multi-release benchmark with known platform attribution and enough English text to compare duplicate-detection methods under a controlled setting. We restrict the analysis to English-language tweets, remove empty text, and retain tweet text, account identifiers, timestamps, language labels, hashtags, URLs, and other available metadata. The final corpus contains 187,000 English-language tweet records from 847 accounts.
\begin{algorithm}[t]
\caption{Generic-Aware Duplicate Campaign Detection}
\label{alg:generic_duplicate_campaigns}
\small
\begin{algorithmic}[1]
\Require English tweets $P$, trained tweet classifier $M$, duplicate methods $\mathcal{D}$
\Ensure Campaign sets and coordination graphs for each method
\State $P' \gets \emptyset$
\For{each tweet $p \in P$}
    \State $t_p \gets$ NormalizeText$(p.text)$
    \If{$t_p$ is not empty}
        \State $e_p \gets$ Embed$(t_p)$
        \State $\hat{y}_p \gets M(e_p)$
        \State Add $(p.account,t_p,e_p,\hat{y}_p)$ to $P'$
    \EndIf
\EndFor
\For{each duplicate method $d \in \mathcal{D}$}
    \State $E_d \gets$ SimilarTweetPairs$(P',d)$
    \State $C_d \gets$ ConnectedComponents$(E_d)$
    \State Remove components with fewer than two accounts from $C_d$
    \For{each campaign $c \in C_d$}
        \State $type(c) \gets$ MajorityLabel of tweet labels in $c$
    \EndFor
    \State $G_{all}^{d} \gets$ BuildGraph$(C_d)$
    \State $G_{ng}^{d} \gets$ BuildGraph$(\{c \in C_d : type(c)=\textsc{Non-Generic}\})$
\EndFor
\State \Return $C_d$, $G_{all}^{d}$, $G_{ng}^{d}$ for all $d \in \mathcal{D}$
\end{algorithmic}
\end{algorithm}
\noindent\textbf{Tweet labeling and classification pipeline.}
We normalize tweet text by removing URLs, mentions, HTML tags, emojis, invisible Unicode characters, and retweet markers; decoding HTML entities; lowercasing; and collapsing whitespace. Hashtag text is retained because it may contain narrative-specific information. We randomly sample 1,000 tweets and label them as \textsc{Generic} or \textsc{Non-Generic} using an LLM-assisted annotation protocol with GPT-5-mini. The model receives only cleaned tweet text, without account identifiers, dataset names, timestamps, or other metadata. The prompt defines generic tweets as common, reusable, vague, low-information, or not tied to a specific claim, event, target, group, policy, or narrative. It defines non-generic tweets as posts containing a specific informational, political, social, or propagandistic message. We then encode the labeled tweets using sentence-transformers/all-mpnet-base-v2 and train supervised classifiers to scale the labels to the remaining English-language tweets.

\begin{table*}[t]
\centering
\scriptsize
\caption{Examples of generic and non-generic duplicate campaign content.}
\label{tab:example_campaigns}
\resizebox{\textwidth}{!}{%
\begin{tabular}{ll}
\toprule
\textbf{Type} & \textbf{Example Tweet} \\
\midrule
\rowcolor{red!15}
Generic & ``A best friend is someone who came into your life, stayed with you through ups and downs, and never left.'' \\

\rowcolor{red!15}
Generic & ``No matter how much you feed the wolf, he keeps looking at the forest.'' \\
\midrule

\rowcolor{green!15}
Non-Generic & ``The Syrian government and Red Crescent helped return families from al-Hawl camp at Albania's request.'' \\

\rowcolor{green!15}
Non-Generic & ``If Cuba were racist apartheid-style dictatorship, no one would be so eager to do business with it'' \\
\bottomrule
\end{tabular}%
}
\end{table*}

\begin{table*}[t]
\centering
\scriptsize
\caption{Average 5-fold cross-validation performance for generic versus non-generic tweet classification.}
\label{tab:classifier_results}
\resizebox{\textwidth}{!}{
\begin{tabular}{
l r
>{\columncolor{red!10}}r
>{\columncolor{red!10}}r
>{\columncolor{red!10}}r
>{\columncolor{green!10}}r
>{\columncolor{green!10}}r
>{\columncolor{green!10}}r
r r
}
\toprule
\multirow{2}{*}{\textbf{Model}}
& \multirow{2}{*}{\textbf{Acc.}}
& \multicolumn{3}{>{\columncolor{red!15}}c}{\textbf{Generic}}
& \multicolumn{3}{>{\columncolor{green!15}}c}{\textbf{Non-Generic}}
& \multirow{2}{*}{\textbf{Macro F1}}
& \multirow{2}{*}{\textbf{Weighted F1}} \\
\cmidrule(lr){3-5} \cmidrule(lr){6-8}
& & \textbf{P} & \textbf{R} & \textbf{F1}
& \textbf{P} & \textbf{R} & \textbf{F1}
& & \\
\midrule
Logistic Regression & 0.911 & 0.767 & \textbf{0.913} & 0.833 & \textbf{0.971} & 0.911 & 0.939 & 0.886 & 0.914 \\
Linear SVM & 0.908 & 0.793 & 0.851 & 0.818 & 0.952 & 0.926 & 0.938 & 0.878 & 0.909 \\
RBF SVM & \textbf{0.926} & 0.828 & 0.884 & \textbf{0.853} & 0.962 & 0.939 & \textbf{0.951} & \textbf{0.902} & \textbf{0.927} \\
Random Forest & 0.885 & \textbf{0.914} & 0.581 & 0.708 & 0.881 & \textbf{0.982} & 0.928 & 0.818 & 0.875 \\
Extra Trees & 0.901 & 0.893 & 0.672 & 0.765 & 0.904 & 0.974 & 0.937 & 0.851 & 0.896 \\
AdaBoost & 0.905 & 0.827 & 0.768 & 0.795 & 0.928 & 0.949 & 0.938 & 0.867 & 0.904 \\
Gradient Boosting & 0.890 & 0.840 & 0.677 & 0.747 & 0.903 & 0.958 & 0.930 & 0.839 & 0.886 \\
KNN & 0.903 & 0.781 & 0.842 & 0.808 & 0.949 & 0.922 & 0.935 & 0.872 & 0.904 \\
MLP Neural Net & 0.920 & 0.852 & 0.818 & 0.831 & 0.943 & 0.953 & 0.948 & 0.889 & 0.919 \\
\bottomrule
\end{tabular}%
}
\end{table*}

\noindent\textbf{Duplicate campaign construction.}
We define a duplicate campaign as a connected component of tweets from at least two distinct accounts that contain identical or highly similar text. To avoid tuning the duplicate-detection procedure to our downstream graph results, we use three previously proposed duplicate-detection settings and keep their parameters fixed across all datasets.

First, Ratcliff/Obershelp serves as a conservative lexical baseline: following prior coordination studies, tweets are sorted chronologically within each dataset and each tweet is compared with the next ten tweets; pairs from different accounts are linked when similarity is at least 0.70~\cite{VNCCGL24,PFM20}. This bounded comparison favors temporally proximate lexical repetition and should therefore be interpreted as a conservative baseline rather than an exhaustive all-pairs search. Second, following prior duplicate-campaign work~\cite{shafin-carbunar-2026-duplicating}, we encode tweets using paraphrase-multilingual-MiniLM-L12-v2 and apply DBSCAN with min\_samples=2 and radius $\epsilon=1.0$. Third, we use sentence-transformers/all-mpnet-base-v2 embeddings with FAISS cosine similarity~\cite{luceri2024unmasking,shafin2026iox}; embeddings are $L_2$-normalized, indexed with inner-product search, and pairs from different accounts are retained when cosine similarity is at least 0.95.

For all methods, linked tweets are merged into connected components, components with fewer than two accounts are removed, and campaign type is assigned by the majority tweet-level label within each component. In our dataset, no campaign contained equal numbers of generic and non-generic tweets, so every campaign received a unique label. Algorithm~\ref{alg:generic_duplicate_campaigns} summarizes the complete pipeline from tweet-level classification to duplicate campaign construction and downstream graph generation.

\noindent\textbf{Graph construction.}
We build weighted account-account graphs by connecting two accounts when they participate in the same duplicate campaign; edge weight increases with the number of shared campaigns. Since a campaign with $n$ accounts can induce up to $n(n-1)/2$ pairwise links, large generic campaigns can disproportionately affect graph structure. For each duplicate-detection method, we construct $G_{all}$ using all campaigns and $G_{ng}$ using only non-generic campaigns, and compare nodes, edges, density, average degree, connected components, largest component size, and average edge weight.

\section{Findings}
\label{sec:findings}

\subsection{Tweet Classification Results}
\label{sec:classification_results}

GPT-5-mini labeled the 1,000 sampled tweets as 241 generic and 759 non-generic. Before training the classifiers, we checked the labeled tweets for exact duplicates after normalization and found that all 1,000 tweets were unique. To assess the reliability of the annotation rubric, two authors independently re-annotated a stratified subset of 200 tweets, consisting of 50 tweets initially labeled as generic and 150 as non-generic by GPT-5-mini. Both annotators used the same generic/non-generic rubric and were blind to the LLM-generated labels and to each other's annotations. They agreed on 194 of the 200 tweets, corresponding to 97.0\% raw agreement and a Cohen's $\kappa$ of 0.919, indicating very high inter-annotator agreement. The annotators agreed on 46 generic and 148 non-generic tweets and disagreed on six tweets. No adjudication was performed before computing inter-annotator agreement. Table~\ref{tab:example_campaigns} presents representative examples of generic and non-generic duplicate content, illustrating the distinction between broadly reusable expressions and context-specific narrative statements.

Table~\ref{tab:classifier_results} reports the average five-fold performance for the supervised classifiers trained on the 1,000 labeled tweets. The RBF SVM performs best, achieving 0.926 accuracy, 0.902 macro-F1, and 0.927 weighted-F1. We use this model to label the remaining English-language tweets before campaign construction. These scores measure agreement with the generic/non-generic annotation rubric, not correctness against a fully human-labeled gold standard or truthfulness judgments.

\subsection{Duplicate Detection Results}
\label{sec:duplicate_detection_results}

\begin{table*}[t]
\centering
\scriptsize
\caption{Campaign-level characteristics across duplicate-detection methods. Duration is measured in days. Hashtag and URL rates are the percentage of campaigns containing at least one tweet with a hashtag or URL.}
\label{tab:campaign_characteristics}
\resizebox{\textwidth}{!}{%
\begin{tabular}{llrrrrrrrrrr}
\toprule
\textbf{Method} & \textbf{Type} & \textbf{Campaigns} & \textbf{Tweets/} & \textbf{Accounts/} & \textbf{Max} & \textbf{Unique} & \textbf{Words/} & \textbf{Avg.} & \textbf{Hashtag} & \textbf{URL} \\
& & & \textbf{Campaign} & \textbf{Campaign} & \textbf{Accounts} & \textbf{Accounts} & \textbf{Text} & \textbf{Duration} & \textbf{\%} & \textbf{\%} \\
\midrule

\multirow{2}{*}{Ratcliff/Obershelp}
& \cellcolor{red!10}Generic
& \cellcolor{red!10}109
& \cellcolor{red!10}2.17
& \cellcolor{red!10}2.13
& \cellcolor{red!10}4
& \cellcolor{red!10}116
& \cellcolor{red!10}5.91
& \cellcolor{red!10}0.12
& \cellcolor{red!10}7.34
& \cellcolor{red!10}33.94 \\

&
\cellcolor{green!10}Non-Generic
& \cellcolor{green!10}2,294
& \cellcolor{green!10}2.80
& \cellcolor{green!10}2.65
& \cellcolor{green!10}33
& \cellcolor{green!10}268
& \cellcolor{green!10}11.09
& \cellcolor{green!10}0.03
& \cellcolor{green!10}11.46
& \cellcolor{green!10}5.27 \\

\midrule

\multirow{2}{*}{Embedding + Clustering}
& \cellcolor{red!10}Generic
& \cellcolor{red!10}2,961
& \cellcolor{red!10}4.28
& \cellcolor{red!10}2.71
& \cellcolor{red!10}21
& \cellcolor{red!10}373
& \cellcolor{red!10}8.59
& \cellcolor{red!10}280.29
& \cellcolor{red!10}4.19
& \cellcolor{red!10}28.54 \\

&
\cellcolor{green!10}Non-Generic
& \cellcolor{green!10}4,690
& \cellcolor{green!10}5.06
& \cellcolor{green!10}4.80
& \cellcolor{green!10}46
& \cellcolor{green!10}348
& \cellcolor{green!10}10.85
& \cellcolor{green!10}17.22
& \cellcolor{green!10}8.91
& \cellcolor{green!10}6.12 \\

\midrule

\multirow{2}{*}{Embedding + Similarity}
& \cellcolor{red!10}Generic
& \cellcolor{red!10}2,949
& \cellcolor{red!10}4.29
& \cellcolor{red!10}2.71
& \cellcolor{red!10}20
& \cellcolor{red!10}375
& \cellcolor{red!10}8.67
& \cellcolor{red!10}273.94
& \cellcolor{red!10}4.27
& \cellcolor{red!10}28.86 \\

&
\cellcolor{green!10}Non-Generic
& \cellcolor{green!10}4,578
& \cellcolor{green!10}5.35
& \cellcolor{green!10}4.89
& \cellcolor{green!10}76
& \cellcolor{green!10}362
& \cellcolor{green!10}11.04
& \cellcolor{green!10}18.01
& \cellcolor{green!10}9.79
& \cellcolor{green!10}6.62 \\

\bottomrule
\end{tabular}%
}
\end{table*}

\begin{table*}[t]
\centering
\scriptsize
\caption{Account-account coordination graph characteristics across duplicate-detection methods and campaign types.}
\label{tab:graph_comparison}
\resizebox{\textwidth}{!}{%
\begin{tabular}{llrrrrrrr}
\toprule
\textbf{Method} & \textbf{Campaign Type} & \textbf{Nodes} & \textbf{Edges} & \textbf{Density} & \textbf{Avg. Degree} & \textbf{Components} & \textbf{Largest Comp.} & \textbf{Avg. Edge Weight} \\
\midrule

\multirow{2}{*}{Ratcliff/Obershelp}
& \cellcolor{red!10}Both Generic and Non-Generic
& \cellcolor{red!10}312
& \cellcolor{red!10}4,086
& \cellcolor{red!10}0.084
& \cellcolor{red!10}26.19
& \cellcolor{red!10}40
& \cellcolor{red!10}138
& \cellcolor{red!10}2.14 \\

&
\cellcolor{green!10}Only Non-Generic
& \cellcolor{green!10}266
& \cellcolor{green!10}4,027
& \cellcolor{green!10}0.114
& \cellcolor{green!10}30.28
& \cellcolor{green!10}30
& \cellcolor{green!10}138
& \cellcolor{green!10}2.14 \\

\midrule

\multirow{2}{*}{Embedding + Clustering}
& \cellcolor{red!10}Both Generic and Non-Generic
& \cellcolor{red!10}497
& \cellcolor{red!10}12,020
& \cellcolor{red!10}0.098
& \cellcolor{red!10}48.37
& \cellcolor{red!10}16
& \cellcolor{red!10}428
& \cellcolor{red!10}8.93 \\

&
\cellcolor{green!10}Only Non-Generic
& \cellcolor{green!10}348
& \cellcolor{green!10}10,433
& \cellcolor{green!10}0.173
& \cellcolor{green!10}59.96
& \cellcolor{green!10}25
& \cellcolor{green!10}166
& \cellcolor{green!10}9.39 \\

\midrule

\multirow{2}{*}{Embedding + Cosine Similarity}
& \cellcolor{red!10}Both Generic and Non-Generic
& \cellcolor{red!10}503
& \cellcolor{red!10}12,096
& \cellcolor{red!10}0.096
& \cellcolor{red!10}48.10
& \cellcolor{red!10}16
& \cellcolor{red!10}435
& \cellcolor{red!10}9.62 \\

&
\cellcolor{green!10}Only Non-Generic
& \cellcolor{green!10}362
& \cellcolor{green!10}10,528
& \cellcolor{green!10}0.161
& \cellcolor{green!10}58.17
& \cellcolor{green!10}26
& \cellcolor{green!10}207
& \cellcolor{green!10}10.17 \\

\bottomrule
\end{tabular}%
}
\end{table*}

The three duplicate-detection methods identify different volumes of duplicate campaigns. Ratcliff/Obershelp detects 2,403 duplicate campaigns, making it the most conservative method. Embedding-based clustering detects 7,651 duplicate campaigns, while embedding-based cosine similarity detects 7,527 duplicate campaigns. Thus, the embedding-based methods recover more than three times as many duplicate campaigns as the lexical baseline.

The generic share also differs sharply across methods. Ratcliff/Obershelp identifies 109 generic campaigns and 2,294 non-generic campaigns, so generic campaigns account for only 4.5\% of its detected campaigns. In contrast, embedding-based clustering identifies 2,961 generic and 4,690 non-generic campaigns, while embedding-based cosine similarity identifies 2,949 generic and 4,578 non-generic campaigns. Generic campaigns therefore account for 38.7\% and 39.2\% of campaigns detected by the two embedding-based methods, respectively. This suggests that semantic duplicate detection is more likely to surface reusable, low-information repetition in addition to narrative-specific duplication.

\subsection{Campaign-Level Differences}
\label{sec:campaign_results}

Table~\ref{tab:campaign_characteristics} compares generic and non-generic campaigns across three detection methods. Across all three methods, non-generic campaigns contain more tweets per campaign and involve more accounts per campaign. Under embedding-based similarity, for example, non-generic campaigns average 5.35 tweets and 4.89 accounts per campaign, compared with 4.29 tweets and 2.71 accounts for generic campaigns.

The textual and temporal characteristics reinforce this distinction. Non-generic campaigns have longer representative text across all methods, suggesting greater narrative specificity. Under embedding-based similarity, non-generic campaigns average 11.04 words per representative text, compared with 8.67 for generic campaigns. Generic campaigns, however, have longer average duration and higher URL prevalence under the embedding-based methods. This suggests that reusable low-information content can recur across broader time spans and may be more strongly associated with URL-sharing behavior, whereas non-generic campaigns are more textually specific and more concentrated in time.

\subsection{Effect on Coordination Graphs}
\label{sec:graph_results}

Table~\ref{tab:graph_comparison} compares account-account coordination graphs constructed from all campaigns with graphs constructed only from non-generic campaigns. For Ratcliff/Obershelp, restricting the graph to non-generic campaigns removes 46 nodes but only 59 edges, and the largest connected component remains unchanged at 138 nodes. This suggests that generic Ratcliff/Obershelp campaigns mostly involve peripheral accounts rather than the main coordination core.

The embedding-based methods show a much stronger effect. With embedding clustering, restricting the graph to non-generic campaigns reduces the largest connected component from 428 to 166 nodes, while density increases from 0.098 to 0.173 and average degree increases from 48.37 to 59.96. The FAISS similarity method shows the same pattern: the largest component decreases from 435 to 207 nodes, while density rises from 0.096 to 0.161. These results indicate that generic duplicate campaigns can increase the apparent scale of duplicate-based coordination networks, especially by connecting many accounts into large components. After filtering to non-generic campaigns, the resulting graphs are smaller and denser, consistent with a more concentrated network around narrative-specific repetition. Because graph density can also increase when peripheral nodes are removed, we interpret this structural change as evidence of concentration rather than, by itself, stronger coordination.

\section{Discussion and Limitations}
\label{sec:discussion}

Our results show that duplicate content should not be treated as uniformly informative evidence of coordination. The effect varies by method. Ratcliff/Obershelp identifies relatively few generic campaigns, while embedding-based methods recover many generic or reusable posts. This broader recall is useful, but it also increases the risk of treating low-information repetition as meaningful narrative coordination.

The long duration of generic campaigns under the embedding-based methods is particularly important. Generic text can be semantically similar across long time spans, causing embedding-based methods to connect posts that are reusable but not necessarily part of the same narrative push. This does not mean such campaigns are irrelevant; generic content may support account maintenance, engagement, or visibility. However, it should be separated from non-generic duplicate campaigns when interpreting coordination graphs.

This study has limitations. First, we analyze only English-language Russian IO datasets. The 39\% generic-campaign rate should therefore not be interpreted as universal; it may vary across languages, platforms, actors, and campaign periods. Our claim is not that this corpus represents all IO behavior, but that generic repetition can materially affect duplicate-based coordination graphs even in a widely used public IO setting. Second, the classifier is trained on LLM-assisted rubric labels rather than a fully human-labeled gold standard. Although two authors independently re-annotated a stratified subset of 200 tweets and achieved 97.0\% raw agreement (Cohen's $\kappa = 0.919$), systematic biases in the LLM-assisted labels could still propagate to the remaining tweets and downstream graph analysis. Third, although the labeled tweets were unique after normalization, future work should evaluate campaign-level or account-level splits to test generalization across near-duplicate groups and accounts. Finally, because this short paper uses duplicate-detection parameters from prior work, we do not conduct a full threshold sensitivity analysis; future work should examine threshold robustness, per-release graph statistics, temporal burstiness, recurrence, and synchronization.

\section{Conclusion}
\label{sec:conclusion}

This paper studies generic duplicate campaigns in duplicate-based coordination analysis of information operations. Using 187,000 English-language tweets from six Russian Twitter Information Operations datasets, we classify tweets as generic or non-generic, construct duplicate campaigns using three methods, and compare coordination graphs built from all campaigns with graphs built only from non-generic campaigns. Generic campaigns account for nearly 39\% of embedding-based duplicate campaigns and substantially affect graph structure. Restricting graphs to non-generic campaigns produces smaller but denser graphs with smaller largest connected components, suggesting that filtering generic campaigns provides a more concentrated view of narrative-specific repetition. These findings highlight a simple but important refinement: repeated content should be evaluated not only by similarity, but also by semantic specificity.

\bibliography{references}

\end{document}